\documentclass[twocolumn,showpacs,superscriptaddress,preprintnumbers,amsmath,amssymb]{revtex4}

\usepackage{graphicx}
\usepackage{dcolumn}
\usepackage{bm}

\usepackage{amssymb}

\begin{document}


\title{Covariant gauge on the lattice: a new implementation}

\author{Attilio Cucchieri}
\affiliation{Instituto de F\'{\i}sica de S\~ao Carlos, Universidade de
             S\~ao Paulo, \\ Caixa Postal 369, 13560-970 S\~ao Carlos, SP, Brazil \\[0.5mm]}
\author{Tereza Mendes}
\affiliation{Instituto de F\'{\i}sica de S\~ao Carlos, Universidade de
             S\~ao Paulo, \\ Caixa Postal 369, 13560-970 S\~ao Carlos, SP, Brazil \\[0.5mm]}
\affiliation{DESY--Zeuthen, Platanenallee 6, 15738 Zeuthen, Germany \\[0.5mm]}
\author{Elton M.\ S.\ Santos}
\affiliation{Instituto de F\'{\i}sica de S\~ao Carlos, Universidade de
             S\~ao Paulo, \\ Caixa Postal 369, 13560-970 S\~ao Carlos, SP, Brazil \\[0.5mm]}
\affiliation{Instituto de Educa\c c\~ao Agricultura e Ambiente, Campus Vale do Rio Madeira,
             Universidade Federal do Amazonas, 69800-000 Humait\'a, AM, Brazil}

\vskip 2mm
\date{\today}

\begin{abstract}
We derive a new implementation of linear covariant gauges on
the lattice, based on a minimizing functional that can be
interpreted as the Hamiltonian of a spin-glass model in
a random external magnetic field.
We show that our method solves most problems encountered in
earlier implementations, mostly related to the no-go condition
formulated by L.\ Giusti, Nucl.\ Phys.\ B 498, 331 (1997).
We carry out tests in the SU(2) case in four space-time dimensions. We also
present preliminary results for the transverse gluon propagator at different
values of the gauge parameter $\xi$.
\end{abstract}

\maketitle


\section{Introduction}

The behavior of Green's functions in the
infrared (IR) limit of Yang-Mills theories is of fundamental importance
for the understanding of the low-energy properties of quantum chromodynamics
(QCD), in particular for the problem of quark and gluon confinement
\cite{Alkofer:2006fu}. Since the evaluation of these functions depends
on the gauge condition, it is important to consider different gauges
in order to obtain a clear (possibly gauge-independent) picture of color
confinement. Needless to say, this investigation should be at the
nonperturbative level.

A nonperturbative study of the QCD propagators and vertices from first
principles is possible using lattice simulations. Of course,
on the lattice, the finite size of the system corresponds to an IR cutoff
$\sim 2 \pi/L$, where $L$ is the lattice size. Thus, a numerical study of
Green's functions in the IR limit usually requires a careful extrapolation
of the data to the infinite-volume limit \cite{constraints}.
Another possible limitation for the simulations is the difficulty in finding 
an efficient numerical implementation of a given gauge condition.
For this reason, most numerical studies of Green's functions have been 
restricted to: Landau gauge \cite{Cucchieri:2008yp}, Coulomb gauge
\cite{Cucchieri:2006hi}, $\lambda$-gauge (a gauge that 
interpolates between Landau and Coulomb) \cite{Cucchieri:2007uj}
and maximally Abelian gauge \cite{MAG}. 
On the other hand, among the various gauge conditions that are very popular 
in continuum studies, the so-called linear covariant gauge --- which is a 
generalization of Landau gauge --- proved quite hostile to the lattice 
approach \cite{Giusti:1996kf,discretization,propagators,Giusti:2000yc,
Cucchieri:2008zx,Mendes:2008ux}.


Let us recall that, in the continuum, the linear covariant gauge is defined by
\begin{equation} 
   \partial_{\mu} A_{\mu}^b(x) \;=\; \Lambda^b(x)
\label{eq:Feynmancontinuo}
\;\mbox{,}
\end{equation}
where $A_{\mu}^b(x)$ is the gluon field and the real-valued functions
$\Lambda^b(x)$ are generated using a Gaussian distribution
\begin{equation}
P\left[\Lambda^b(x)\right] \;\sim\; exp{\Bigl\{ - \, \sum_b\,
\left[ \Lambda^b(x) \right]^2 / \, (2 \, \xi) \Bigr\} } \;\mbox{.}
\label{eq:gaussian}
\end{equation}
The Feynman gauge corresponds to the value $\xi = 1$, while
the Landau gauge is obtained in the limit $\xi \to 0$.

In the Landau case, the gauge condition
$ \partial_{\mu} A_{\mu}^b(x) \,=\, 0 \, $
can be obtained by minimizing the functional
\begin{equation}
  {\cal E}_{LG}\{A^{g}\} \; \propto \;
     \int \, d^4x \; \sum_{\mu, b} \left[ (A^{g})_{\mu}^b(x) \right]^2
\label{eq:Landau}
\end{equation}
with respect to the gauge transformations $\{g(x)\}$. In Ref.\ \cite{Giusti:1996kf}
it was shown that a similar minimizing functional ${\cal E}_{LCG}\{A^{g}\}$ for the
linear covariant gauge does not exist.

The no-go theorem proven in \cite{Giusti:1996kf} can of course be avoided
by relaxing its hypotheses. The first possibility, explored in that
reference, is to consider a different gauge, i.e.\ $ F [ \, \partial_{\mu}
A_{\mu}^b(x) \,-\, \Lambda^b(x) \, ] \, = \, 0 $
with $F[0]=0$, for which a minimizing functional exists. In
particular it was shown that minimizing the functional
\begin{equation}
      \int \, d^4x \;\sum_{\mu, b} \left\{
   \left[ \partial_{\mu} A_{\mu}^b(x) - \Lambda^b(x) \right]^2
             \right\}
\label{eq:giusti}
\end{equation}
implies the stationarity condition
\begin{equation}
D_{\nu}^{ab} \partial_{\nu} \left[ \partial_{\mu} A_{\mu}^b(x)
        - \Lambda^b(x) \right] \, = \, 0 \; ,
\end{equation}
where $\, D_{\nu}^{ab} \,$ is the covariant derivative.

As noted previously \cite{Giusti:1996kf}, this method presents several problems.
First of all, one can introduce spurious solutions, corresponding to $F[s]=0$
for $s \neq 0$. In the above case, these solutions are the zeros of the operator
$\, D_{\nu}^{ab} \partial_{\nu} \, $. Also, the second derivative
of this functional does not correspond to the Faddeev-Popov operator
${\cal M}\,=\,- \partial_{\mu} D_{\mu}^{ab}$ of the usual linear covariant gauge.
Finally, the lattice discretization of the above functional is not linear in
the gauge transformation $\{ g(x) \}$.
This makes the numerical minimization difficult and one has to rely on a specific
discretization of the minimizing functional \cite{discretization}
in order to make the lattice approach feasible.

A second possibility, recently presented in \cite{Cucchieri:2008zx}, is based on
avoiding the use of a minimizing functional ${\cal E}_{LCG}\{A^{g}\}$, i.e.\ on
considering a lattice definition of the linear covariant gauge that coincides
with the perturbative definition in the continuum. In this case, one
first fixes the gluon field to Landau gauge,
i.e.\ the transformed gauge field satisfies $\partial_{\mu} A_{\mu}^b(x)
= 0$. Then, one solves the equation $ \left( \partial_{\mu} D_{\mu}^{bc}
\phi^c\right)(x) = \Lambda^b(x)$
and uses $\phi^c(x)$ as a generator of a second gauge transformation.
For small $ \phi^c(x)$,
one then has that the gauge-transformed gluon field ${A'}_{\mu}^{b}(x)$
satisfies the condition
\begin{equation}
\partial_{\mu} {A'}_{\mu}^{b}(x) = \partial_{\mu}
    \left(A_\mu^b + D_\mu^{bc}\phi^c\right)(x) = \Lambda^b(x) \; .
\end{equation}

The main problem in this case is that the method is correct only for
infinitesimal gauge transformations, but usually $\phi^c(x)$ is not small
in a numerical simulation. As a consequence, one finds \cite{Cucchieri:2008zx}
that the distribution of $\partial_{\mu} {A'}_\mu^{b}(x)$ does not agree
completely with the Gaussian distribution of $\Lambda^b(x)$. Moreover,
in linear covariant gauge the longitudinal gluon propagator $D_l(p^2)$
should satisfy the relation $p^2 D_l(p^2) = \xi$. With this approach one
finds that, for small momenta, this is not the case \cite{Cucchieri:2008zx}.


Here we present a new implementation of the linear covariant gauge on 
the lattice that solves
the problems illustrated above, afflicting earlier methods. The paper is
organized as follows. In the next section, we describe our new implementation
and we report some tests of the algorithm used for the numerical gauge fixing.
In particular, we show that $D_l(p^2)$ is well described by $\xi / p^2$, where
$\xi$ is the gauge parameter, and discuss discretization effects.
We also present preliminary results for the momentum-space transverse gluon
propagator $D_t(p^2)$ for different values of $\xi$. Finally, we present our
conclusions.


\section{A new implementation}
\label{sec:new}

In order to find a new implementation for the linear covariant gauge on the
lattice we notice that, when minimizing a functional
${\cal E}\{A^{g}\}$, the gauge condition is given
by the first variation of ${\cal E}$, i.e.
\begin{equation}
\delta {\cal E} \, = \,
\frac{\partial {\cal E}}{\partial A} \,
 \frac{\partial A}{\partial g} \, \delta g \, = \, 0 \; .
\end{equation}
Ref.\ \cite{Giusti:1996kf} has proven that there is no functional ${\cal E}\{A^{g}\}$ 
leading to Eq.\ (\ref{eq:Feynmancontinuo}). Nevertheless, one can remove an implicit
hypothesis of the no-go condition, i.e.\ that the gauge transformation $\{ g(x) \}$
appears in the minimizing functional in the ``canonical'' way
$A^{g}$. Thus, we may look for a minimizing functional
of the type ${\cal E}_{LCG}\{A^{g}, g\}$ instead of simply
${\cal E}_{LCG}\{A^{g}\}$.

If one recalls that solving the system of equations $B \psi = \zeta$
is equivalent to minimizing the quadratic form $ \psi B \psi / 2 \,
- \, \psi \, \zeta $, then it is clear that we should have
$ \, {\cal E}_{LCG}\{A^{g}, g\} \, \sim \, {\cal E}_{LG}\{A^{g}\} \,
- \, g \Lambda \,$, where
\begin{eqnarray}
\!\!\!\!\!\!\!\!\!{\cal E}_{LG}\{U^{g}\} \!\! & = & \! \!
              - \; \Re \; Tr \sum_{x, \mu}
   \; g(x) \, U_{\mu}(x) \, g^{\dagger}(x+ e_{\mu})
          \label{eq:ELandau2} 
\end{eqnarray}
is the minimizing functional for the lattice Landau gauge.
Here, the link variables $U_{\mu}(x)$ and the
site variable $g(x)$ are matrices belonging to the SU($N_{c}$) group
(in the fundamental representation). We also indicate with $\Re\,$ the
real part of a complex number and with $Tr$ the trace in color space. Indeed,
the lattice linear covariant gauge condition can be obtained by minimizing 
the functional
\begin{equation}
 {\cal E}_{LCG}\{U^{g}, g\} \; = \; {\cal E}_{LG}\{U^{g}\}
                   \, + \, \Re \; Tr \sum_x
                     \,  i\, g(x) \, \Lambda(x) \; .
\label{eq:EFeynman}
\end{equation}
To prove that this is the right functional we can consider a one-parameter
subgroup $ g(x,\tau) \, = \, \exp \left[ i \tau \gamma^{b}(x) \lambda^{b} \right] $
of the gauge transformation $\{ g(x) \}$. Here, $\, \lambda^{b} $
are the traceless Hermitian generators of the Lie algebra of the
SU($N_{c}$) gauge group. They also satisfy the usual normalization condition
$ Tr \left( \lambda^{b} \lambda^{c} \right) \, = \, 2 \, \delta^{bc} $. Then, it is
easy to check that the stationarity condition
$\partial_{\tau} {\cal E}_{LCG} (\tau =0) = 0 \, $ [for all $\gamma^{b}(x)$]
implies the lattice linear covariant gauge condition
\begin{equation}
 \nabla \cdot A^{b}(x) \, = \, \sum_{\mu} \, A_{\mu}^{b}(x)
                        \, - \, A_{\mu}^{b}(x-e_{\mu}) \, = \,
   \Lambda^{b}(x)\; .
   \label{eq:Feynman}
\end{equation}
Here, $ \Lambda^{b}(x) = Tr \, \Lambda(x) \, \lambda^{b} $
and, similarly, $ \, A_{\mu}^{b}(x) = Tr \, A_{\mu}(x) \, \lambda^{b} $
with
\begin{equation}
A_{\mu}(x) \, = \, (2i)^{-1} \, \left[ U_{\mu}(x) - U_{\mu}^{\dagger}(x)
                    \right]_{traceless} \; .
\label{eq:discret}
\end{equation}
Note that Eq.\ (\ref{eq:Feynman}) implies that $ \sum_x \Lambda^{b}(x) = 0$.

At the same time, one can verify that the second term on the r.h.s.\
of Eq.\ (\ref{eq:EFeynman})
does not contribute to the second variation
of the functional ${\cal E}_{LCG}\{U^{g}, g\}$
with respect to the parameter $\tau$.
Thus, this second variation defines a matrix $\, {\cal M} \,$ that
is a discretized version of the usual Faddeev-Popov operator
$- \partial \cdot D$. Let us recall that, due to the gauge condition
(\ref{eq:Feynmancontinuo}) above, in linear covariant gauge one has
in general $ - \partial \cdot D \neq - D \cdot \partial $.

Finally, let us note that the functional ${\cal E}_{LCG}\{U^{g}, g\}$
in Eq.\ (\ref{eq:EFeynman}) is indeed linear in the gauge
transformation $\{ g(x) \}$. Also, one can
interpret the Landau-gauge functional $ {\cal E}_{LG}\{U^{g}\} $
[see Eq.\ (\ref{eq:ELandau2}) above] as a {\em spin-glass}
Hamiltonian for the {\em spin} variables $g(x)$ with a {\em random} interaction
given by $U_{\mu}(x)$. Then, our new functional $ {\cal E}_{LCG}\{U^{g}, g\}$
corresponds to the same spin-glass Hamiltonian when a random external {\em magnetic
field} $ \Lambda(x) $ is applied.


\begin{figure}[t]
\vskip -14.8mm
\hskip -2.5mm
\includegraphics[scale=0.31]{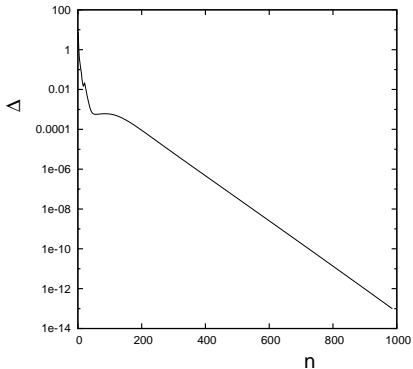}
\vskip -10mm
\caption{\label{fig:gfix}
  Convergence of the numerical gauge fixing. Here we report the value of
  $ \Delta = \sum_{x,b} [ \nabla \cdot A^{b}(x) - \Lambda^{b}(x) ]^2 $ as
  a function of the number of iterations $n$ for $\beta = 4$, $\xi = 0.5$ and $V = 8^4$.
  Note the logarithmic scale on the $y$ axis.
  }
\end{figure}

We have performed some numerical tests with the functional (\ref{eq:EFeynman}),
using the so-called {\em stochastic-overrelaxation} algorithm \cite{gfixing}.
To this end we considered the 4d $SU(2)$ case at $\beta = 4$, for $V = 8^4$ and $16^4$,
with $\xi = 0.001, 0.1$ and 0.5. The numerical gauge fixing seems to work very well
in these cases (see plot in Figure \ref{fig:gfix}).

We also checked that the quantity $p^2 D_l(p^2)$ is constant within
statistical fluctuations in all cases considered. For $V = 16^4$
and $\xi = 0.5$ a fit of the type $a/p^b$ for $D_l(p^2)$ gives $a = 0.502(5)$
and $b = 2.01(1)$ with a $\chi^2/dof = 1.2$. Similar fits have been obtained in
the other cases.


\subsection{Discretization effects}

The minimizing functional (\ref{eq:EFeynman}) in principle solves the problem of fixing
the gauge condition (\ref{eq:Feynman}). From the numerical point of view, however,
one has to recall that the gluon field is bounded, at least when using the
standard (compact) discretization. Thus, while our method works wery well 
when the functions $\Lambda^b(x)$ are generated using a bounded distribution, 
care must be taken in the usual implementation of the linear covariant 
gauge, where the functions $\Lambda^b(x)$ [see Eqs.\
(\ref{eq:Feynmancontinuo}) and (\ref{eq:gaussian})] satisfy a Gaussian
distribution, i.e.\ they are unbounded. This can give rise to convergence
problems \cite{Rank} when a numerical implementation is attempted. Of course
the problem gets more severe when $\xi$ is larger.
This is a common problem of all lattice realizations of the linear 
covariant gauge.

Actually, in order to obtain the correct continuum limit \cite{Cucchieri:2008zx},
the functions $\Lambda^b(x)$ are generated on the lattice, in the SU($N_c$) case,
from a Gaussian distribution with width $\sqrt{\sigma} = \sqrt{2 N_c \xi/\beta}$,
instead of the width $\sqrt{\xi}$. Note that, for $\beta = 4$ and $N_c = 2$,
as in the previous section, one has $\sigma = \xi$. Thus, for a given value of
$\xi$ one can obtain a sufficiently small value for $\sigma$ by considering large
values of the lattice coupling $\beta$. However, if $\beta$ is too large the
physical volume will be too small (for a given lattice size) and one cannot
really probe the IR limit of the theory. On the other hand, for $\beta < 2 N_c$
the lattice width $\sqrt{\sigma}$ is even larger than the continuum width $\sqrt{\xi}$.
Note that the situation is probably better in the $SU(3)$ case, since one has
$\sigma =  \xi$ for $\beta = 6$, which corresponds to a reasonably large value
of the lattice spacing $a$.

Of course, one can try to use different discretizations of the gluon fields
in order to improve the convergence of the minimizing algorithms. Besides the
usual discretization (\ref{eq:discret}), we also did some tests with the ``angle''
projection \cite{Amemiya:1998jz} and with the recently introduced
stereographic projection (or modified lattice Landau gauge) \cite{stereographic},
using the so-called {\em Cornell} method \cite{gfixing}.
In the last case the gluon field is in principle unbounded even for a finite
lattice spacing. Indeed, our tests show that with this discretization one is usually
able to simulate at slightly larger values of $\xi$, for a given lattice volume
$V$ and lattice coupling $\beta$, compared to the other two cases. In particular,
we tested these three discretizations using $V=8^4$, $\xi=0.01,
0.05,0.1,0.5,1.0$ and $\beta=2.2,2.3,\ldots,2.9,3.0$. We found that, while
the usual discretization and the ``angle'' projection have problems with
$\xi \geq 0.5$ already at $\beta = 2.9$, the stereographic projection allows one to
simulate with $\xi=1$ for $\beta \geq 2.5$. Note that $\xi=1$ and $\beta = 2.5$
corresponds to $\sigma = 1.6$.


\begin{figure}[t]
\vskip -12mm
\includegraphics[scale=0.29]{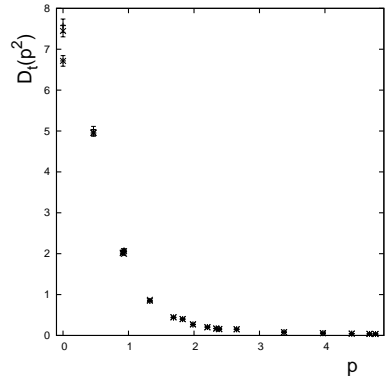}
\vskip -10mm
\caption{\label{fig:Dt16}
    Transverse gluon propagator $D_t(p^2)$ as a function of the
    momentum $p$ (both in physical units) for the lattice volume
    $V = 16^4$, $\beta = 2.3$ and
    $\xi = 0 \, (+), 0.05 \, (\times)$, $0.1 \, (*)$. 
  }
\vskip -2mm
\end{figure}

\subsection{Transverse gluon propagator}
\label{sec:results}

\begin{figure}[t]
\vskip -12mm
\includegraphics[scale=0.29]{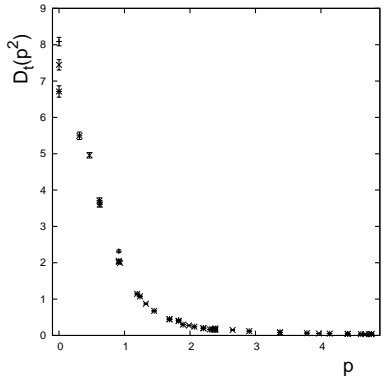}
\vskip -10mm
\caption{\label{fig:Dt}
    Transverse gluon propagator $D_t(p^2)$ as a function of the
    momentum $p$ (both in physical units) for the gauge coupling $\xi = 0.05$,
    $\beta = 2.3$, with the lattice volumes
    $V = 8^4 \, (+), 16^4 \, (\times)$ and $24^4 \, (*)$.
  }
\end{figure}

Using the minimizing functional shown above and the stereographic projection,
we have simulated at $\beta = 2.2$ and $\beta = 2.3$ for the lattice volumes
$V = 8^4, 16^4$ and $24^4$, for several values of the gauge parameter $\xi$
in the SU(2) case. As a test of the gauge-fixing method, we have checked that
the quantity $D_l(p^2) p^2/\sigma$, which should be equal to 1, has a value of
$0.999(2)$ when averaged over all data $D_l(p^2)$ produced. Preliminary results
for the transverse gluon propagator $D_t(p^2)$ as a function of the momentum
$p$ are shown in the Figures \ref{fig:Dt16} and \ref{fig:Dt}. There is a clear
tendency of getting a more suppressed IR propagator when the lattice volumes
increases, as in the Landau case, and also when the value of $\xi$ increases.
The latter result is in agreement with Ref.\ \cite{Giusti:2000yc}. Also, the
extrapolation to infinite volume, for a given $\beta$ and a fixed value of $\xi$,
seems in this case even harder than in Landau gauge. Indeed, as $V \to \infty$,
the number of sites characterized by a large value for the function
$\Lambda^b(x)$ increases, making the convergence of the gauge-fixing method
more difficult.

One should recall here that, at the perturbative level, the gluon
field $A$ and the gauge parameter $\xi$ are (multiplicatively)
renormalized by the same factor $Z_3$, i.e.\ $A_B = Z_3^{1/2} A_R$
and $\xi_B = Z_3 \, \xi_R$, where as usual $B$ and $R$ indicate bare and
renormalized quantities respectively. On the lattice this implies that,
in the scaling region, data obtained for two different values of $\beta$,
e.g.\ $\beta_1$ and $\beta_2$, can be compared and should give the same
(renormalized) propagator only if the multiplicative factor $R_Z =
Z_3(\beta_1) / Z_3(\beta_2)$ relating the propagators (see
\cite{Leinweber:1998uu} for the case of Landau gauge) also relates
the gauge parameters $\xi_1$ and $\xi_2$. Since the value of $R_Z$ is not
known a priori, one has to use some type of ``matching procedure''
in order to find pairs of parameters $(\beta, \xi)$ yielding the same
continuum renormalized propagators. This is left to a future work
\cite{preparation}.


\section{Conclusions}
\label{sec:concl}

We presented a new lattice implementation of the linear covariant gauge
by using a minimizing functional ${\cal E}_{LCG}\{U^{g}, g\}$. Tests have been
done for the SU(2) case in four space-time dimensions. This approach
solves most problems encountered in earlier implementations and ensures
a good quality for the gauge fixing with a ratio $D_l(p^2) p^2/\xi \approx 1$
for all cases considered. We have also reported preliminary results for the transverse
gluon propagator $D_t(p^2)$. The only open problem is how to extend these
simulations to large lattice volumes, in order to probe the IR limit of the theory
when the gauge parameter $\xi$ is also large. 
We stress that infinite-volume results for $D_t(p^2)$ would of course be 
very important for comparison
with analytic studies \cite{analytic,Sobreiro:2005vn} in the continuum. 
In any case, as mentioned above, the discretization effects are
probably less severe for the SU(3) group compared to the SU(2) case. 
We are currently
simulating other values of $\beta$ and $\xi$ in the 4d SU(2) case and
considering simulations also of the SU(3) group and of the 3d case \cite{preparation}.

Finally, having a minimizing functional for the linear covariant gauge, which
extends in a natural way the Landau case while preserving all the properties of the
continuum formulation, allows a numerical investigation of the effects of Gribov
copies on Green's functions for the case $\xi \neq 0$. These studies will make
possible a comparison to recent analytic results \cite{Sobreiro:2005vn} obtained
in the limit of small gauge parameter $\xi$. They could also be important for
understanding how the so-called (Landau) Gribov-Zwanwiger confinement scenario
\cite{Cucchieri:2008yp} should be modified in the general covariant gauge. 
To this end, results obtained for small values of $\xi$, which can be easily
obtained using our new implementation, could already be relevant.


\vskip 2mm

A.C. and T.M. acknowledge partial support from FAPESP
and from CNPq. The work of T.M. was supported also by
the Alexander von Humboldt Foundation.
E.M.S.S. acknowledges support from CAPES.




\begin{thebibliography}{99}

\bibitem{Alkofer:2006fu}
  R.\ Alkofer and J.\ Greensite,
  J.\ Phys.\ G {\bf 34}, S3 (2007). 

\bibitem{constraints}
  A.\ Cucchieri and T.\ Mendes,
  Phys.\ Rev.\ Lett.\ {\bf 100}, 241601 (2008); 
  Phys.\ Rev.\ D {\bf 78}, 094503 (2008). 

\bibitem{Cucchieri:2008yp}
  For a review see e.g.\ A.\ Cucchieri and T.\ Mendes,
  arXiv:0809.2777 [hep-lat].

\bibitem{Cucchieri:2006hi}
  For a review see e.g.\ A.\ Cucchieri,
  AIP Conf.\ Proc.\ {\bf 892}, 22 (2007) [arXiv:hep-lat/0612004].

\bibitem{Cucchieri:2007uj}
  A.\ Cucchieri, A.\ Maas and T.\ Mendes,
  Mod.\ Phys.\ Lett.\ A {\bf 22}, 2429 (2007). 

\bibitem{MAG}
  V.G.\ Bornyakov {\em et al.},
  Phys.\ Lett.\ B {\bf 559}, 214 (2003); 
  T.\ Mendes, A.\ Cucchieri and A.\ Mihara,
  AIP Conf.\ Proc.\ {\bf 892}, 203 (2007)
  [arXiv:hep-lat/0611002].

\bibitem{Giusti:1996kf}
  L.\ Giusti,
  Nucl.\ Phys.\ B {\bf 498}, 331 (1997). 

\bibitem{discretization}
  L.\ Giusti, M.L.\ Paciello, S.\ Petrarca and B.\ Taglienti,
  Nucl.\ Phys.\ Proc.\ Suppl.\ {\bf 83}, 819 (2000); 
  Phys.\ Rev.\ D {\bf 63}, 014501 (2001). 

\bibitem{propagators}
  L.\ Giusti, M.L.\ Paciello, S.\ Petrarca and B.\ Taglienti,
  arXiv:hep-lat/9912036;
  L.\ Giusti {\it et al.}, 
  Nucl.\ Phys.\ Proc.\ Suppl.\ {\bf 106-107}, 995 (2002). 

\bibitem{Giusti:2000yc}
  L.\ Giusti {\it et al.}, 
  Nucl.\ Phys.\ Proc.\ Suppl.\ {\bf 94}, 805 (2001). 

\bibitem{Cucchieri:2008zx}
  A.\ Cucchieri, A.\ Maas and T.\ Mendes,
  Comput.\ Phys.\ Commun.\ {\bf 180}, 215 (2009). 

\bibitem{Mendes:2008ux}
  T.\ Mendes, A.\ Cucchieri, A.\ Maas and A.\ Mihara,
  arXiv:0809.3741 [hep-lat].

\bibitem{gfixing}
  A.\ Cucchieri and T.\ Mendes,
  Nucl.\ Phys.\ B {\bf 471}, 263 (1996); 
  Nucl.\ Phys.\ Proc.\ Suppl.\ {\bf 53}, 811 (1997); 
  Comput.\ Phys.\ Commun.\ {\bf 154}, 1 (2003). 

\bibitem{Rank} J.\ Rank, {\em Thermal Screening Masses in the Standard Model of
               Strong and Electroweak Interactions}, Ph.D. thesis,
               Bielefeld University, January 1998,
               http://www. physik.uni-bielefeld.de/theory/e6/publiframe.html.

\bibitem{Amemiya:1998jz}
  K.\ Amemiya and H.\ Suganuma,
  Phys.\ Rev.\ D {\bf 60}, 114509 (1999). 

\bibitem{stereographic}
  L.\ von Smekal, D.\ Mehta, A.\ Sternbeck and A.G.\ Williams,
  PoS {\bf LAT2007}, 382 (2007). 
  A.\ Sternbeck and L.\ von Smekal,
  arXiv:0811.4300 [hep-lat].

\bibitem{Leinweber:1998uu}
  D.B.\ Leinweber, J.I.\ Skullerud, A.G.\ Williams and C.\ Parrinello
  [UKQCD Collaboration],
  Phys.\ Rev.\ D {\bf 60}, 094507 (1999)
  [Erratum-ibid.\ D {\bf 61}, 079901 (2000)]. %

\bibitem{preparation}
    A.\ Cucchieri, T.\ Mendes, G.\ Nakamura and E.\ M.\ S.\ Santos, in preparation.

\bibitem{analytic}
  R.\ Alkofer, C.S.\ Fischer, H.\ Reinhardt and L.\ von Smekal,
  Phys.\ Rev.\ D {\bf 68}, 045003 (2003);
  A.C.\ Aguilar and J.\ Papavassiliou,
  Phys.\ Rev.\ D {\bf 77}, 125022 (2008).

\bibitem{Sobreiro:2005vn}
  R.F.\ Sobreiro and S.P.\ Sorella, JHEP {\bf 0506}, 054 (2005).

\end{thebibliography}
\end{document}